\documentclass[11pt,twoside]{article}

\usepackage{asp2006}
\usepackage{epsf}
\usepackage{lscape}
\usepackage{comment}

\begin{document}
\title{Balancing the Energy Budget: Star-Formation versus AGN in High Redshift Infrared Luminous Galaxies}   
\author{E.J.~Murphy,\altaffilmark{1} R.-R. Chary,\altaffilmark{1} 
D.M.~Alexander,\altaffilmark{2} 
M.~Dickinson,\altaffilmark{3} 
B.~Magnelli,\altaffilmark{4} 
G.~Morrison,\altaffilmark{5,6} 
A.~Pope\altaffilmark{3}, H. I. Teplitz\altaffilmark{1}}
\altaffiltext{1}{\scriptsize {\it Spitzer} Science Center, MC 314-6, Caltech, Pasadena, CA 91125; emurphy@ipac.caltech.edu}
\altaffiltext{2}{\scriptsize Department of Physics, Durham University, Durham DH1 3LE, UK}
\altaffiltext{3}{\scriptsize National Optical Astronomy Observatory, Tucson, AZ 85719}
\altaffiltext{4}{\scriptsize Laboratoire AIM, CEA/DSM-CNRS-Universit\'{e} Paris Diderot, France}
\altaffiltext{5}{\scriptsize  Institute for Astronomy, University of Hawaii, Honolulu, HI 96822}
\altaffiltext{6}{\scriptsize Canada-France-Hawaii Telescope, Kamuela, HI 96743}

\begin{abstract} 
We present deep {\it Spitzer} mid-infrared spectroscopy, along with 16, 24, 70, and 850\,$\micron$\ photometry, for 22 galaxies located in the Great Observatories Origins Deep Survey-North (GOODS-N) field.  
The sample spans a redshift range of $0.6\la z \la 2.6$, 24~$\mu$m flux densities between $\sim$0.2$-$1.2 mJy, and consists of submillimeter galaxies (SMGs), X-ray or optically selected active galactic nuclei (AGN), and optically faint ($z_{AB}>25$\,mag) sources. 
We find that infrared (IR; $8-1000~\micron$) luminosities derived by fitting local spectral energy distributions (SEDs) with 24~$\micron$ photometry alone are well matched to those when additional mid-infrared spectroscopic and longer wavelength photometric data is used for galaxies having $z\la1.4$ and 24~$\micron$-derived IR luminosities typically $\la 3\times 10^{12}~L_{\sun}$.  
However, for galaxies in the redshift range between $1.4\la z \la 2.6$, typically having 24~$\micron$-derived IR luminosities $\ga 3\times 10^{12}~L_{\sun}$, IR luminosities are overestimated by an average factor of $\sim$5 when SED fitting with 24~$\micron$ photometry alone.  
This result arises partly due to the fact that high redshift galaxies exhibit aromatic feature equivalent widths that are large compared to local galaxies of similar luminosities.  
Through a spectral decomposition of mid-infrared spectroscopic data, we are able to isolate the fraction of IR luminosity arising from an AGN as opposed to star formation activity.   
This fraction is only able to account for $\sim$30\% of the total IR luminosity among the entire sample.  
\end{abstract}



\vspace{-1cm}
\section{Introduction}
The mid-infrared ($5-40~\micron$) SED is a complex interplay of broad emission features thought to arise from polycyclic aromatic hydrocarbon (PAH) molecules, silicate absorption features at 9.7 and 18~$\micron$, and a mid-infrared continuum from very small grains \citep{lp84,atb85}.
Mid-infrared luminosities measured near $\sim$8~$\micron$ have been found to correlate with the total infrared (IR; $8-1000$) luminosities of galaxies in the local Universe \citep{ce01, de02}, which itself is a measure of a galaxy's SFR.  
While a correlation is found, there does exist a large amount of scatter \citep{dd05,jd07,la07} and systematic departures for low metallicity systems \citep{ce05,sm06}.  
It is therefore uncertain if these empirical relations between mid-infrared and total IR luminosities ($L_{\rm IR}$) for galaxies in the local Universe are applicable for galaxies at higher redshifts.  

Recently, data from deep far-infrared surveys such as the Far-infrared Deep Legacy Survey (FIDEL; PI: M. Dickinson) are confirming the same redshift evolution of the SFR density seen in the mid-infrared and do not show any evolution in the SED of infrared luminous galaxies \citep[e.g.][]{bm09}.
In contrast, \citet{jr08} have provided evidence that the high redshift lensed mid-infrared selected galaxies might show a factor of $\sim$2 stronger rest-frame 8~$\micron$ emission compared to their total IR luminosities.

Using mid-infrared spectroscopy for 22 sources selected at 24~$\micron$ in the GOODS-N field, along with existing 850~$\micron$ and additional 70~$\micron$ imagery obtained as part of FIDEL, 
we aim to improve our understanding of the star formation and AGN activity within a diverse group of $0.6\la z \la2.6$ galaxies.  
This is done through a more proper estimate of IR luminosities and the ability to decompose these measurements into star-forming and AGN components.  
For the full study, please see \citet{ejm09}.  

\section{Accuracy of 24~$\mu$m-Derived IR Luminosities \label{sec-fitcomp}}
\begin{figure}
{\resizebox{13cm}{!}{
\plottwo{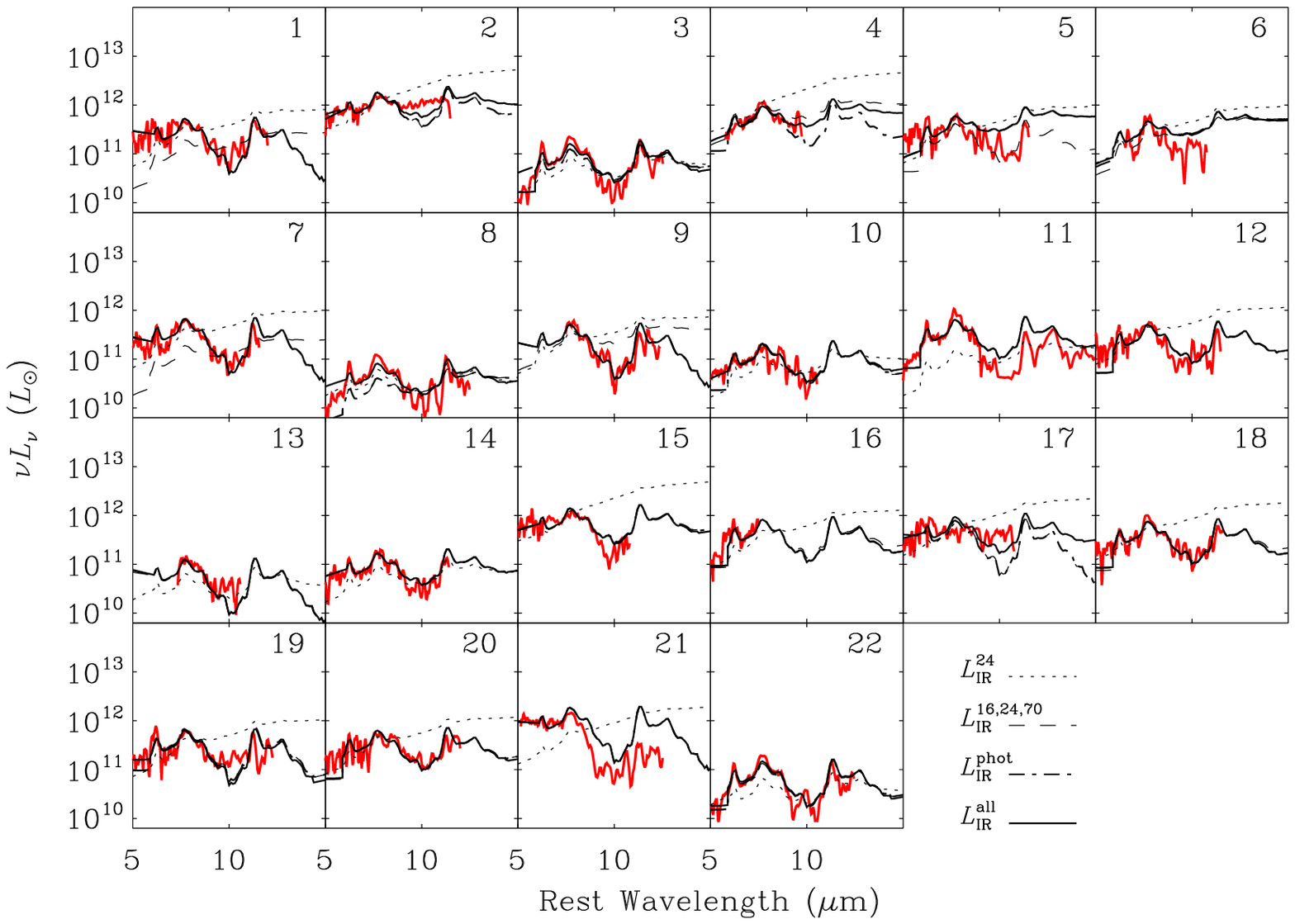}{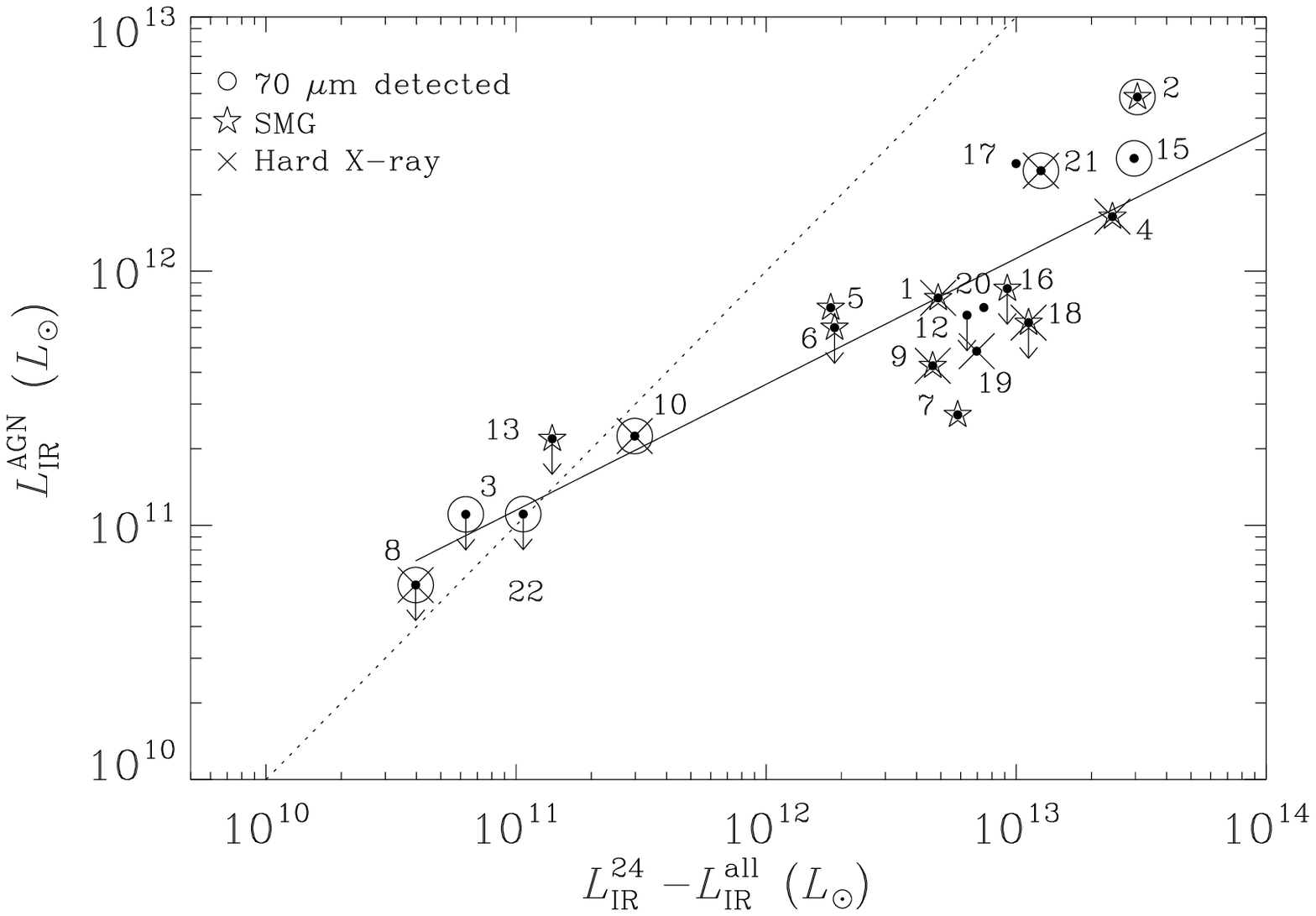}}}
\caption{\scriptsize 
{\it Left}: A blowup of the $5-15~\micron$ range for the best-fit SEDs of \citet{ce01} for all sources using only 24~$\micron$ flux densities ($\it dotted$ lines), all 3 {\it Spitzer} (16, 24, and 70~$\micron$) flux densities ({\it dashed} lines), {\it Spitzer} and submillimeter (850~$\micron$) photometry ({dot-dashed} lines), and all photometry ({\it Spitzer} and submillimeter) along with the IRS spectra ({\it solid} lines).  
The actual spectra, smoothed to the instrumental resolution, are overplotted in red.  
The SEDs chosen when fitting with the 24~$\micron$ data alone rarely characterize the observed PAH emission as 
compared to when longer wavelength data is included in the fitting.    
{\it Right}: The estimated AGN luminosity as a function of the difference between the 24~$\micron$ and best-fit IR luminosities.
Sources for which only upper limits to the AGN mid-infrared luminosity fraction could be determined are shown as {\it downward arrows}.  
An ordinary least squares fit is overplotted as a {\it solid} line while the {\it dotted} line indicates a one-to-one relationship to show that the AGN luminosity is not able to account for the difference in the bolometric correction when fitting the SED templates with 24~$\micron$ photometry alone.     
\label{fig-1}}
\end{figure}

By plotting the $5-15~\micron$ mid-infrared regions of the four different \citet{ce01} fits (see caption) for each galaxy in the left panel of Figure \ref{fig-1}, one easily sees that by using 24~$\micron$ flux densities alone the equivalent widths of the aromatic features observed in the IRS spectra are 
underestimated by the \citet{ce01} templates, resulting in highly discrepant estimates (i.e. overestimates) of IR luminosity.  

The range of IR luminosities derived by fitting the 24~$\micron$ photometry alone span a range between $4.3\times10^{11}-4.2\times10^{13}~L_{\sun}$ (i.e. nearly two orders of magnitude) with a median of $7.5\times10^{12}~L_{\sun}$.  
Looking at the best-fit estimates for the IR luminosities among the entire sample, we find that they range between $3.2\times10^{11}-1.1\times10^{13}~L_{\sun}$ (i.e. a factor of $\sim$35) having a median of $2.2\times10^{12}~L_{\sun}$.  
The 24~$\micron$-derived IR luminosity is a factor of $\sim$4 larger than that for the best-fit estimates, on average.  

Using our best-fit IR luminosites, we also note that 16 out of the 22 sample galaxies are classified as ULIRGs while the remaining 6 are LIRGs.  
Looking at the differences between the 24~$\micron$ to our best-fit IR luminosities, the discrepancy is found to be much larger among the ULIRGs than for the LIRGs.  
The 24~$\micron$-derived IR luminosities are larger than the best-fit IR luminosities by a factor of $\sim$5, on average, for the ULIRGs while being only $\sim$1.2 times larger, on average, for the LIRGs.

\section{SFR Estimates and the Contribution from AGN}
In the right panel of Figure \ref{fig-1} we plot AGN luminosity versus the difference between the 24~$\micron$ and best-fit estimate IR luminosities.  
A clear trend of increasing AGN luminosity with increasing overestimation of the IR luminosity using 24~$\micron$ photometry alone is observed;  or, in other words, AGN luminosity appears to scale with increasing mid-infrared luminosity.  
Performing an ordinary least squares fit to the data, we find that 
\begin{equation}
\label{eq-agnfrac}
\log\left(\frac{L_{\rm IR}^{\rm AGN}}{L_{\sun}}\right) = 
	(0.50 \pm 0.04) \log \left(\frac{L_{\rm IR}^{24} - L_{\rm IR}^{\rm all}}{L_{\sun}}\right) + (5.61 \pm 0.52).
\end{equation}
While the differences between the 24~$\micron$ and best-fit IR luminosities are large for galaxies having large (i.e. $\ga$60\%) mid-infrared AGN fractions, we find these quantities to be unrelated for galaxies having smaller mid-infrared AGN fractions.  
We also find that the AGN luminosity accounts for only 16\% of the difference between the 24~$\micron$ derived and best-fit IR luminosities, on average.    
This indicates that the AGN contribution to the mid-infrared excesses, which would result in overestimates of IR-based SFRs, is close to negligible when compared to the improper bolometric correction applied when estimating total IR luminosities by fitting local SED templates with 24~$\micron$ data alone.  
SFRs derived by subtracting the AGN fraction from the 24~$\micron$ photometry are larger than the AGN corrected best-fit IR SFRs by a factor of $\sim$1 to $\sim$3.3, on average for redshifts below and above $z\sim1.4$, respectively.  
Furthermore, the AGN-corrected 24~$\micron$ SFRs are factors of $\sim$1 to $\sim$68 times larger than the corresponding UV-corrected SFRs for sample galaxies at redshifts above $z\sim1.4$.  
Therefore, the AGN is not the dominant source of the inferred ``mid-infrared excesses" \citep[see][]{ed07a} among these systems.  

\section{Conclusions}
In the present study we have used observations from the mid-infrared to the submillimeter to properly characterize the IR luminosities for a diverse sample of 22 galaxies spanning a redshift range of $0.6 \la z \la 2.6$.  
In addition, we have used the mid-infrared spectra of these sources to estimate the fractions of their IR luminosities which arise from an AGN.  
Our conclusions can be summarized as follows:

\begin{enumerate}

\item
IR ($8-1000~\micron$) luminosities derived by SED fitting observed 24~$\micron$ flux densities alone are well matched to those when additional mid-infrared spectroscopy and 16, 70, and 850~$\micron$ photometry are included in the fits for galaxies having $z \la 1.4$ and $L_{\rm IR}^{\rm 24}$ values typically  $\la$$3\times10^{12}~L_{\sun}$.  
In contrast, for galaxies lying in a redshift range between 1.4 and 2.6 with $L_{\rm IR}^{\rm 24}$ values typically $\ga$$3\times10^{12}~L_{\sun}$, IR luminosities derived by SED template fitting using observed 24~$\micron$ flux densities alone overestimate the true IR luminosity by a factor of $\sim$5, on average, compared to fitting all available data.  
A comparison between the observed mid-infrared spectra with that of the SEDs chosen from fitting 24~$\micron$ photometry alone and from fitting all available photometric data demonstrates that local high luminosity SED templates show weaker PAH emission by an average factor of $\sim$5 in this redshift range and do not properly characterize the contribution from PAH emission.   

\item
After decomposing the IR luminosity into star forming and AGN components, we find the AGN luminosity to be increasing with increasing difference between the 24~$\micron$-derived and our best-fit IR luminosities.  
Such a trend suggests that the AGN power increases with mid-infrared luminosity.    
However, we also find that the median fraction of the AGN to the difference between the 24~$\micron$-derived and best-fit IR luminosities is only 16\% suggesting the AGN power is almost negligible compared to the bolometric correction necessary to properly calibrate the 24~$\micron$-derived IR luminosities.  

\end{enumerate}


\end{document}